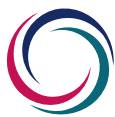

# Single-molecule conductance of a chemically modified, π-extended tetrathiafulvalene and its charge-transfer complex with F$_4$TCNQ

Raúl García[1], M. Ángeles Herranz[1], Edmund Leary[*,2], M. Teresa González[2], Gabino Rubio Bollinger[3], Marius Bürkle[4], Linda A. Zotti[5], Yoshihiro Asai[4], Fabian Pauly[6], Juan Carlos Cuevas[5], Nicolás Agraït[2,3] and Nazario Martín[*,1,2]

## Full Research Paper

Open Access

Address:
[1]Departamento de Química Orgánica, Facultad de Química, Universidad Complutense, E-28040 Madrid, Spain, [2]Fundación IMDEA Nanoscience, Campus de Cantoblanco, Universidad Autónoma, E-28048 Madrid, Spain, [3]Departamento de Física de la Materia Condensada, and Instituto "Nicolás Cabrera", Universidad Autonoma de Madrid, E-28049 Madrid, Spain, [4]Nanosystem Research Institute, National Institute of Advanced Industrial Science and Technology (AIST), Tsukuba, Ibaraki 305-8568, Japan, [5]Departamento de Física Teórica de la Materia Condensada, Universidad Autónoma de Madrid, Spain and [6]Department of Physics, University of Konstanz, D-78457 Konstanz, Germany

Email:
Edmund Leary[*] - edmund.leary@imdea.org; Nazario Martín[*] - nazmar@quim.ucm.es

* Corresponding author

## Abstract

We describe the synthesis and single-molecule electrical transport properties of a molecular wire containing a π-extended tetrathiafulvalene (exTTF) group and its charge-transfer complex with F$_4$TCNQ. We form single-molecule junctions using the in situ break junction technique using a homebuilt scanning tunneling microscope with a range of conductance between 10 G$_0$ down to $10^{-7}$ G$_0$. Within this range we do not observe a clear conductance signature of the neutral parent molecule, suggesting either that its conductance is too low or that it does not form a stable junction. Conversely, we do find a clear conductance signature in the experiments carried out on the charge-transfer complex. Due to the fact we expected this species to have a higher conductance than the neutral molecule, we believe this supports the idea that the conductance of the neutral molecule is very low, below our measurement sensitivity. This idea is further supported by theoretical calculations. To the best of our knowledge, these are the first reported single-molecule conductance measurements on a molecular charge-transfer species.

## Introduction

The development of molecular electronics is a current challenge in nanoscience. The ultimate goal is to fabricate different electronic devices based on a variety of active elements such as wires, transistors, diodes or switches (to name just a few), where each is built from individual, suitably functionalized molecules. Although the controlled handling of molecules to form reliable molecule-based circuits remains a demanding task, modern organic synthesis allows the design and preparation of nearly any challenging molecule [1].

Among the aforementioned electroactive elements, the study of molecular wires has received great attention and, in this regard, a great variety of molecules of different nature involving single, double and triple C–C bonds (conjugated or not) have been extensively studied [2-4]. Furthermore, most of these systems have been covalently connected to a great variety of different anchor groups in order to improve the connection to various metal electrodes. However, despite the huge number of molecular systems that find application as wires, the use of electroactive molecules exhibiting different oxidation states that can modify/control the conductance through the wire have been considerably less studied.

In the realm of organic chemistry there are a great variety of organic compounds able to show different redox states that are very appealing candidates to be used as wires and/or switches in molecular electronics. In this regard, molecules having redox centers such as viologen [5,6], aniline [7,8], thiophene [9], anthraquinone [10] and ferrocene [11] have been previously studied. However, a particularly suitable redox-active molecule for molecular electronics is the well-known electron donor tetrathiafulvalene (TTF) molecule. Pristine TTF, as well as the tetraselenafulvalene analogue (TSF), have been previously reported. In this study, the authors hypothesized that in the Au–TTF–Au junctions, the molecule is connected to the electrodes in a face-to-face overlapping configuration [12]. In contrast, since the first report on a suitably functionalized TTF as a molecular wire using two thioacetate anchoring groups [13], most of the TTF derivatives synthesized for this purpose have been functionalized with sulfur atoms as alligator clips. These belong either to a fused ring on the TTF [14], or to a chain covalently connected to the TTF core bearing a thiol group at the end [15,16]. Furthermore, extended TTF cruciform molecules, formed by two orthogonally placed, π-conjugated moieties bearing the 1,3-dithiole rings at the ends, have also been used for single-molecule measurements [17].

A singular TTF analogue is the so-called π-extended TTF (exTTF, (9,10-bis(1,3-dithiol-2-ylidene)-9,10-dihydroanthracene) which, in contrast to pristine TTF that exhibits two oxi-dation peaks to form the radical cation and dication species, shows only one oxidation peak involving a two electron process to form the dication state. Furthermore, the geometrical properties of exTTF are quite different from pristine TTF. Thus, whereas TTF is mostly planar in the neutral form, exTTF is a highly distorted, out-of-plane molecule with a butterfly shape in its neutral state. It undergoes a dramatic gain of planarity and aromaticity upon oxidation (Figure 1) [18,19]. This gain of stability upon oxidation has been skillfully used in a variety of D–π–A systems, namely exTTF–π-bridge–$C_{60}$ derivatives, for determining the attenuation factor of the molecular wire (oligomer) acting as the π-bridge [20-22].

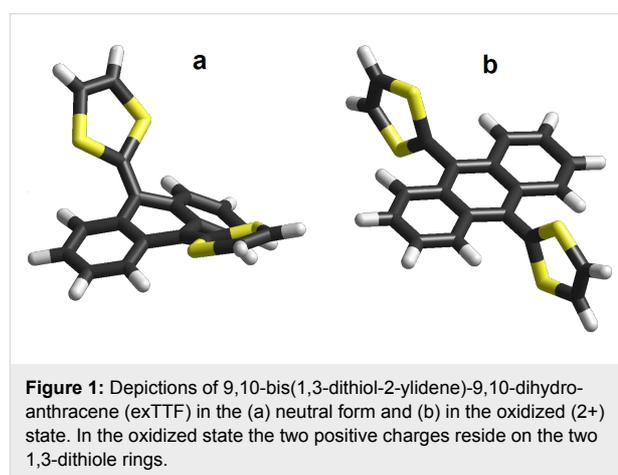

**Figure 1:** Depictions of 9,10-bis(1,3-dithiol-2-ylidene)-9,10-dihydroanthracene (exTTF) in the (a) neutral form and (b) in the oxidized (2+) state. In the oxidized state the two positive charges reside on the two 1,3-dithiole rings.

In this paper, we describe the synthesis of a new exTTF derivative, suitably functionalized with two (*p*-acetylthio)phenylethynyl substituents at positions 2 and 6 of the anthracene central core. We then describe single-molecule conductance measurements on this new derivative, along with measurements of the charge-transfer complex formed with $F_4TCNQ$. Finally, we present theoretical calculations to understand its electrical transport properties.

## Results and Discussion
### Synthesis and characterization of molecular wire 5

The synthesis of the target molecule is shown in Scheme 1 and starts from the previously reported 2,6-diiodo-exTTF **1** [23]. Reaction of **1** under Sonogashira conditions (Pd(II), CuI, DIPEA) with trimethylsilylacetylene affords the symmetrically substituted exTTF **2** in good yield. Further removal of the trimethylsilyl group is easily achieved by treatment with potassium carbonate, yielding the free terminal alkyne **3** in quantitative yield. The introduction of the two anchor groups in **3** is carried out through a second Sonogashira reaction. The coupling of 2,6-diethynyl-exTTF **3** with 1-acetylthio-4-iodoben-zene (**4**), in the presence of Pd(II) and copper iodide and DIPEA in THF, led to the target molecule **5** in moderate yield (35%). Compound **4** was obtained in one step and with high yield from 1,4-diiodobenzene by reaction with *n*-BuLi in ether, followed by reaction with $S_8$, and eventually, acetyl chloride [24].

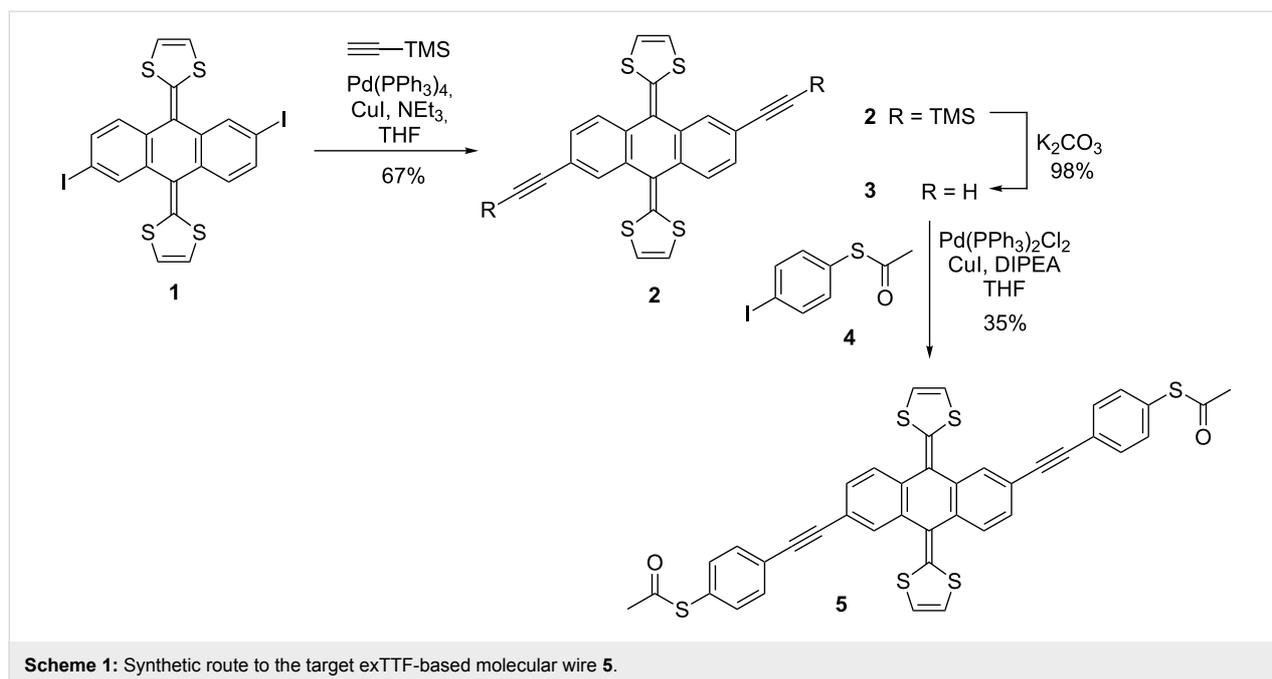

**Scheme 1:** Synthetic route to the target exTTF-based molecular wire **5**.

All new compounds were fully characterized by spectroscopic and electrochemical means. Interestingly, compounds **2**, **3** and **5** showed similar $^1$H NMR spectra due to their symmetry resulting in relatively simple spectra, which confirms the proposed structures. In particular, compound **5** exhibits the methyl groups as a singlet at 2.45 ppm, and the protons corresponding to the 1,3-dithiole rings appear as a singlet at 6.37 ppm. In the $^{13}$C NMR of the target molecule **5**, the carbonyl groups appear at 192.4 ppm and the carbons of the alkyne moieties at 88.0 and 90.4 ppm, with the terminal methyl groups at 28.7 ppm.

The redox properties of compound **5** (0.2 mM) were determined by cyclic voltammetry at room temperature in THF using $TBAPF_6$ (0.1 M) as a supporting electrolyte under argon atmosphere and at a scan rate of 0.1 Vs$^{-1}$. The electrochemical cell consisted of a glassy carbon working electrode, Ag/AgNO$_3$ reference electrode and a Pt wire counter electrode. It is worth mentioning that ferrocene was not employed as the inner reference since its oxidation potential overlaps with that of the exTTF unit. Similar to pristine exTTF, the new exTTF derivative **5** exhibited only one quasi-reversible oxidation peak, involving a two-electron process to form the dication state [25,26]. This oxidation peak appears at $E_{pa}$ = 217 mV ($\Delta E$ = 285 mV, peak-to-peak separation), which is quite similar to the oxidation peak found for pristine exTTF ($E_{pa}$ = 244 mV, $\Delta E$ = 350 mV) (Figure 2).

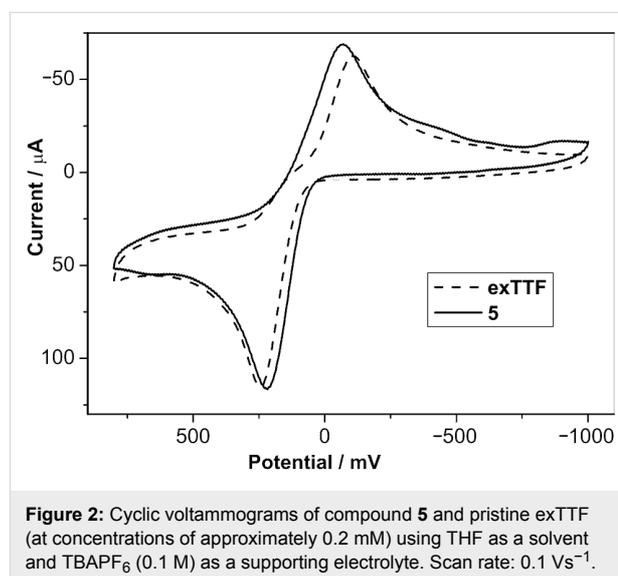

**Figure 2:** Cyclic voltammograms of compound **5** and pristine exTTF (at concentrations of approximately 0.2 mM) using THF as a solvent and $TBAPF_6$ (0.1 M) as a supporting electrolyte. Scan rate: 0.1 Vs$^{-1}$.

## Break junction measurements
### Neutral molecule

We first tried to form neutral-molecule molecular junctions. Simply, compound **5** can be seen as an analogue of an oligo(phenylene ethynylene), specifically an OPE3-dithiol compound (where 3 indicates the total number of phenyl rings), in which the central phenyl ring has been substituted by an exTTF unit. We could, therefore, expect compound **5** to form molecular junctions in a similar way to the OPE3-dithiol, which readily does so either in solution, or under solvent-free conditions as recently reported [27]. Hence, as a starting point, we followed the same sample preparation conditions as previously with the OPE3-dithiol [27]. We prepared 0.1–1 mM solutions of compound **5** in both 1,2,4-trichlorobenzene (TCB) or mesitylene/tetrahydrofuran (Mes/THF 4:1), and exposed a clean gold substrate to the solution for a period of approximately 30 min. The sample was then dried under a flow of argon and mounted inside the scanning tunneling microscope (STM). All experiments were then performed under solvent-free, ambient conditions. In order to form molecular junctions of compound **5**, we followed the break junction technique [28]. During the experiment, the variation in conductance ($G$) is recorded while an STM tip is moved vertically ($z$) in and out of contact with a gold substrate, forming and breaking gold nanocontacts ($G$ vs $z$ trace). When the two gold electrodes (the STM tip and the substrate) are in contact, or in close enough proximity, one or more molecules of the compound adsorbed on the surface can bridge the two electrodes. This binding occurs through one of the terminal groups (thiols in this case) to each of the electrodes, hence forming a molecular junction. In this configuration, when separating the electrodes (larger $z$ values), we observe conductance plateaus while the gold nanocontact or a molecular junction remains intact during the pulling. The plateaus then end abruptly when the junction is broken.

Figure 3c and 3d show the 2D histograms consisting of several thousand individual $G$ vs $z$ traces recorded in a break junction experiment in the presence of compound **5**. These are compared with those recorded on an unmodified gold substrate and in the presence of OPE3-dithiol. These histograms include all the measured traces without filtering out the tunneling-only traces from those with plateaus. One can see that the colored region of the 2D histogram of OPE3-dithiol extends significantly to larger inter-electrode distances than the unmodified gold substrate. Also, it does so within a narrow band of conductance values (roughly between $\log(G/G_0) = -3$ and $-4.5$). This 2D profile exemplifies the presence of plateaus occurring in a narrow region of conductance. In the 2D histograms of compound **5**, we can also see that the colored region extends to higher electrode interdistance values ($z$) than for the unmodified gold substrate. However, as opposed to the case of OPE-dithiol, there is no clear protuberance in the histogram that would indicate the concentration of plateaus at a given conductance value.

Therefore, in order to determine whether the signature of compound **5** is simply very weak, we performed a trace separation process to build histograms of only the traces displaying plateaus. Figure 4 shows the results of this separation for the traces considered in Figure 3. In particular, for compound **5**, two separation steps were carried out, resulting in three cate-gories of traces: (1) those with only a smooth exponential decay (Figure 4a and Figure 4d, labeled "No Plat."), (2) those with poorly-defined plateaus (Figure 4b and Figure 4e, labeled "Plat. 1"), and (3) those with well-defined plateaus (Figure 4c and Figure 4f, labeled "Plat. 2"). We considered the conductance range between $\log(G/G_0) = -5.3$ and $-0.5$ and established that a trace has a plateau whenever a $z$ interval ($\Delta z$) longer than 0.15 nm is needed to observe a change of conductance $\Delta \log(G/G_0) = 0.12$ along the trace. Amongst these traces, we separate the well-defined plateaus as those traces for which a $\Delta z$ larger than 0.2 nm is needed to observe a $\Delta \log(G/G_0) = 0.1$ along the trace. These are empirical criteria which gave us the best separation results. We see that the histograms built from the trace of the third group present protuberances, suggesting that molecular junctions are successfully formed from compound **5**.

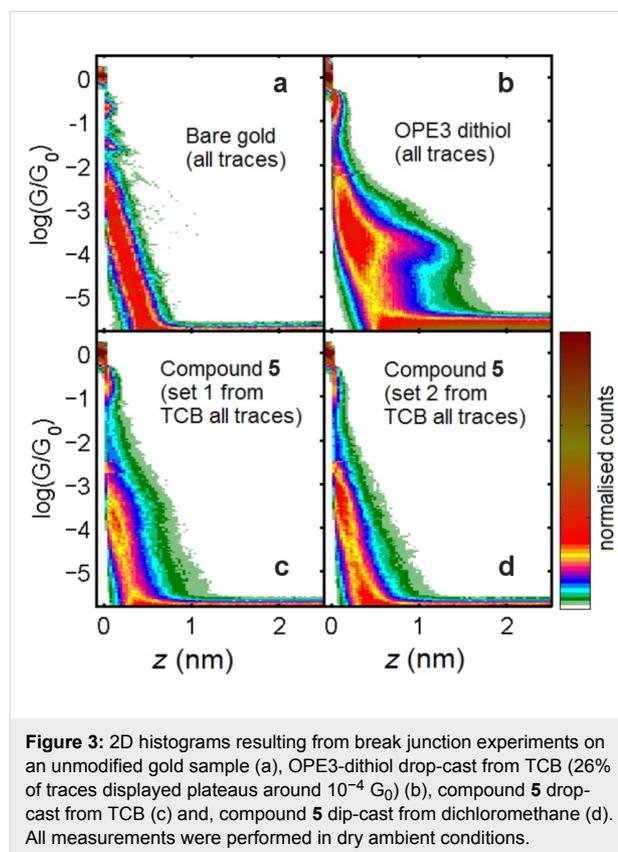

**Figure 3:** 2D histograms resulting from break junction experiments on an unmodified gold sample (a), OPE3-dithiol drop-cast from TCB (26% of traces displayed plateaus around $10^{-4}$ $G_0$) (b), compound **5** drop-cast from TCB (c) and, compound **5** dip-cast from dichloromethane (d). All measurements were performed in dry ambient conditions.

From a close inspection of Figure 4c and Figure 4f, we see that the plateau conductance in these experiments varies by more than 2 orders of magnitude (between $\log(G/G_0) = -3$ and $-5$), significantly greater than OPE3-dithiol, and also that they extend to between 1 and 2 nm. However, it is important to note here that the percentage of traces with well-defined plateaus (given in parentheses in Figure 4) is very low. Even with a total number of 5000 measured traces, the final number of traces with well-defined plateaus is less than 50 traces. We also note here that these results correspond to the most successful measurement runs, and the percentage of traces with plateaus was even lower in other cases. We stress that under the same experimental conditions, we obtained percentages of around 35% of well-defined plateaus for OPE3-dithiol.

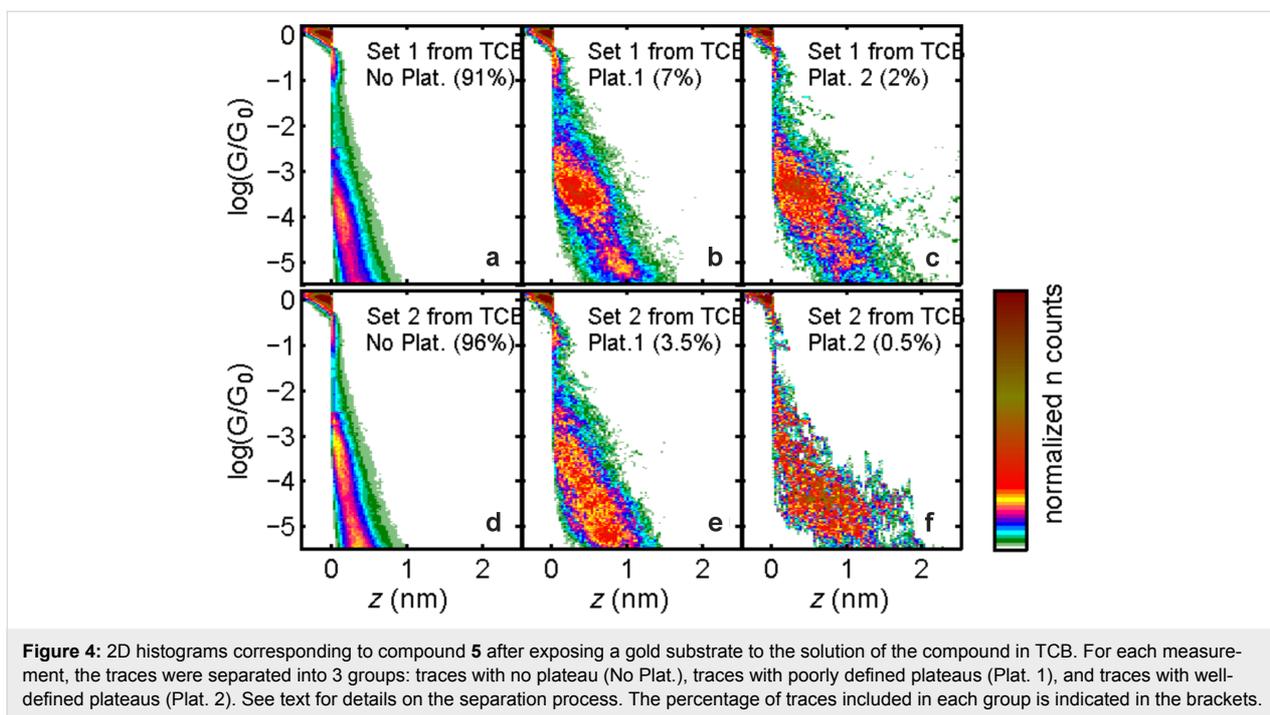

**Figure 4:** 2D histograms corresponding to compound **5** after exposing a gold substrate to the solution of the compound in TCB. For each measurement, the traces were separated into 3 groups: traces with no plateau (No Plat.), traces with poorly defined plateaus (Plat. 1), and traces with well-defined plateaus (Plat. 2). See text for details on the separation process. The percentage of traces included in each group is indicated in the brackets.

We have recently measured the conductance of a series of molecules based on a similar OPE3 backbone, also terminated with thiols, but with differing numbers of dithiafulvalene (DTF) substituents placed at various positions along the OPE backbone [17]. One of the main structural differences with these molecules compared to compound **5** is that conjugation is maintained along the OPE3 backbone. For this series of compounds, we could observe a clear signature for each, which was similar to the unmodified OPE3. We showed that the presence of the DTF side groups does not influence the low-bias conductance and secondly, and that their presence does not significantly hinder molecular junction formation. Different to the measurements carried out thus far on compound **5**, we carried out these measurements using dichloromethane as the deposition solvent, and since the results were clear, we decided to applied this also to compound **5**. The sample was prepared by exposing the gold to a solution of compound **5** in dichloromethane (DCM) for 1 h, followed by drying the sample with a flow of $N_2$. The results of these measurements are shown in Figure 5.

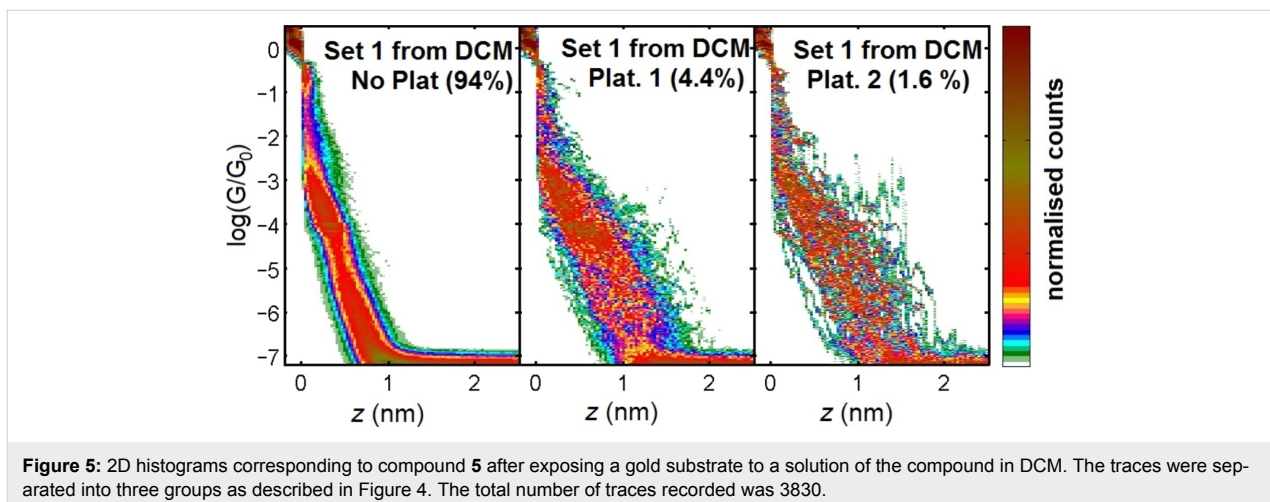

**Figure 5:** 2D histograms corresponding to compound **5** after exposing a gold substrate to a solution of the compound in DCM. The traces were separated into three groups as described in Figure 4. The total number of traces recorded was 3830.

As with the measurements on compound **5** using the TCB drop-casting method, after dip-casting from DCM, we could observe the presence of G(z) plateaus, in this case down to just above $10^{-7}$ G$_0$ facilitated by the use of larger gains. The percentage of traces displaying these plateaus was again low, however, and similar to that found before, with less than 2% fitting the well-defined criteria. From the 2D histograms in Figure 4, we see again that no favored region of conductance develops after separating the junctions displaying plateaus (labeled Plat. 1 and Plat. 2). This, we believe, indicates that whilst the formation of molecular junctions is possible, it is unclear exactly how the molecule binds in these junctions. The low percentage of molecular junction formation can be considered consistent with our observations on the OPE3 series with DTF side groups, in that the presence of sulfur atoms in the center of the molecule slightly reduces the probability of forming junctions. This is likely because the mobility of the molecule over the surface is reduced, which in turn makes it difficult for molecules to diffuse into the freshly created nanogaps.

The lack of a clear molecular signature may ultimately be due to several reasons. Aside from reducing the probability of junction formation, the sulfur atoms in the center of the molecule can also bind to the electrodes during the evolution of the junctions, preventing the wire from becoming fully stretched. This is a distinct possibility for this molecule due to the well-known interaction of the parent extended-TTF and gold [29]. An alternative explanation for the lack of a clear signal for compound **5** is that its end-to-end conductance is too low for it to be observed in our setup. If this is the case, this would mean the conductance is lower than $10^{-7}$ G$_0$. This is quite likely due to the cross-conjugated nature of the exTTF unit, and the known effect this has on conductance.

## Conductance measurements on the exTTF-F$_4$TCNQ charge-transfer complex

Despite the lack of a clear signal for the neutral molecule, we decided to proceed by trying to form the charge-transfer (CT) complex of compound **5** with 7,7,8,8-tetracyanoquinodimethane (TCNQ). When combining two equimolar solutions of the two compounds in DCM, we observed no color change, suggesting no complexation. The solvent was then changed to acetonitrile, a more polar solvent better able to stabilize complex formation, but this too did not give the anticipated strong color change. Only when the solution was heated at reflux for 3 h under ambient conditions and a large excess of TCNQ was added (5 equiv), a dark green color developed. UV–vis absorption spectroscopy confirmed the formation of a CT complex (Supporting Information File 1, Figure S1). To ensure the quantitative formation of the CT complex, a more straightforward method compatible with the conductance measurements was considered. Specifically, we switched from TCNQ to the stronger acceptor 2,3,5,6-tetrafluoro-7,7,8,8-tetracyanoquinodimethane (F$_4$TCNQ). This produced a clear change of color as soon as the first drops of the acceptor were added to the solution of compound **5** in DCM, yielding a green solution. We monitored the formation of the CT complex by recording the UV–vis spectrum of the solution. As successive amounts of the acceptor are added to the solution of the donor, the peak at 448 nm of the neutral donor species decreases and new peaks appear between 600–900 nm that grow with the addition of more acceptor and also methanol (see Supporting Information File 1, Figure S2). Specifically, we observe three peaks at 689 nm, 764 nm and 867 nm, which can be assigned to the radical anion F$_4$TCNQ species [30]. For the break junction measurements, it was necessary to avoid having an excess of the acceptor because F$_4$TCNQ forms molecular junctions itself due to its four terminal cyano groups (see Supporting Information File 1, Figure S7). Hence, we formed the charge-transfer (CT) complex by adding approximately 0.5 mL of a $10^{-4}$ M solution of F$_4$TCNQ in DCM to a 1 mL $10^{-4}$ M solution of compound **5**. In this way, we can avoid having significant amounts of free acceptor on the surface. This, however, means we do not exactly know the ratio of donor to acceptor in our case. It is known that the CT complex of the parent exTTF with TCNQ forms in a 1:4 ratio [31]. However, as we have evidence of the formation of a radical anion of F$_4$TCNQ, we believe a donor to acceptor ratio of 1:2 is more likely, for which there is a precedent in a substituted exTTF complex with TCNQ [32]. We then allowed 24 h for the molecules to adsorb onto the gold in order to obtain as high a density of molecules as possible. This increases the possibility that a significant fraction of the solution is still the free donor species. Although there will still be some of the neutral donor species present on the surface, as we have shown that this molecule does not give a clear signal in break junction experiments, there will be no problem of signal overlap with the CT complex.

In contrast to the neutral form of compound **5**, a significant percentage of conductance plateaus for the CT complex sample were observed. They were found fall into two main groups, labeled as high and low conductance. Figure 6a shows examples of individual G(z) traces displaying conductance plateaus. Firstly, we separate traces showing only tunneling (Figure 6b) from those containing plateaus (Figure 6c) using the following criteria: z interval (Δz) longer than 0.12 nm needed to observe a change of conductance $\Delta\log(G/G_0) = 0.16$. We then further divided the traces into two more groups for those with plateaus above or below $\log(G/G_0) = -3.8$. Dividing the traces using a value slightly above or below this does not change the separation significantly as the difference between the two types of plateaus is clear. By fitting a Gaussian function to the histograms we find peak positions of $\log(G/G_0) = -3.0$ for the high group, and $\log(G/G_0) = -4.7$ for the low group (see Supporting Information File 1, Figure S6 for the 1D histograms). We also measured the junction break-off distance for the two groups (which we define as the separation required to move from $\log(G/G_0) = -0.5$ to either $\log(G/G_0) = -4$ for the high plateaus, or $\log(G/G_0) = -6.1$ for the low plateaus). We found mean values of $0.65 \pm 0.25$ nm and $1.14 \pm 0.30$ nm for the high and low groups, respectively (see Supporting Information File 1, Figure S3 for break-off histograms). We repeated the measurements using a freshly prepared CT complex and gold electrodes, and obtained a very similar result for the low conductance group (see Supporting Information File 1, Figures S4 and S5). In the repeated measurement, however, we did not observe a signal in the high conductance region, above $10^{-3}$ $G_0$.

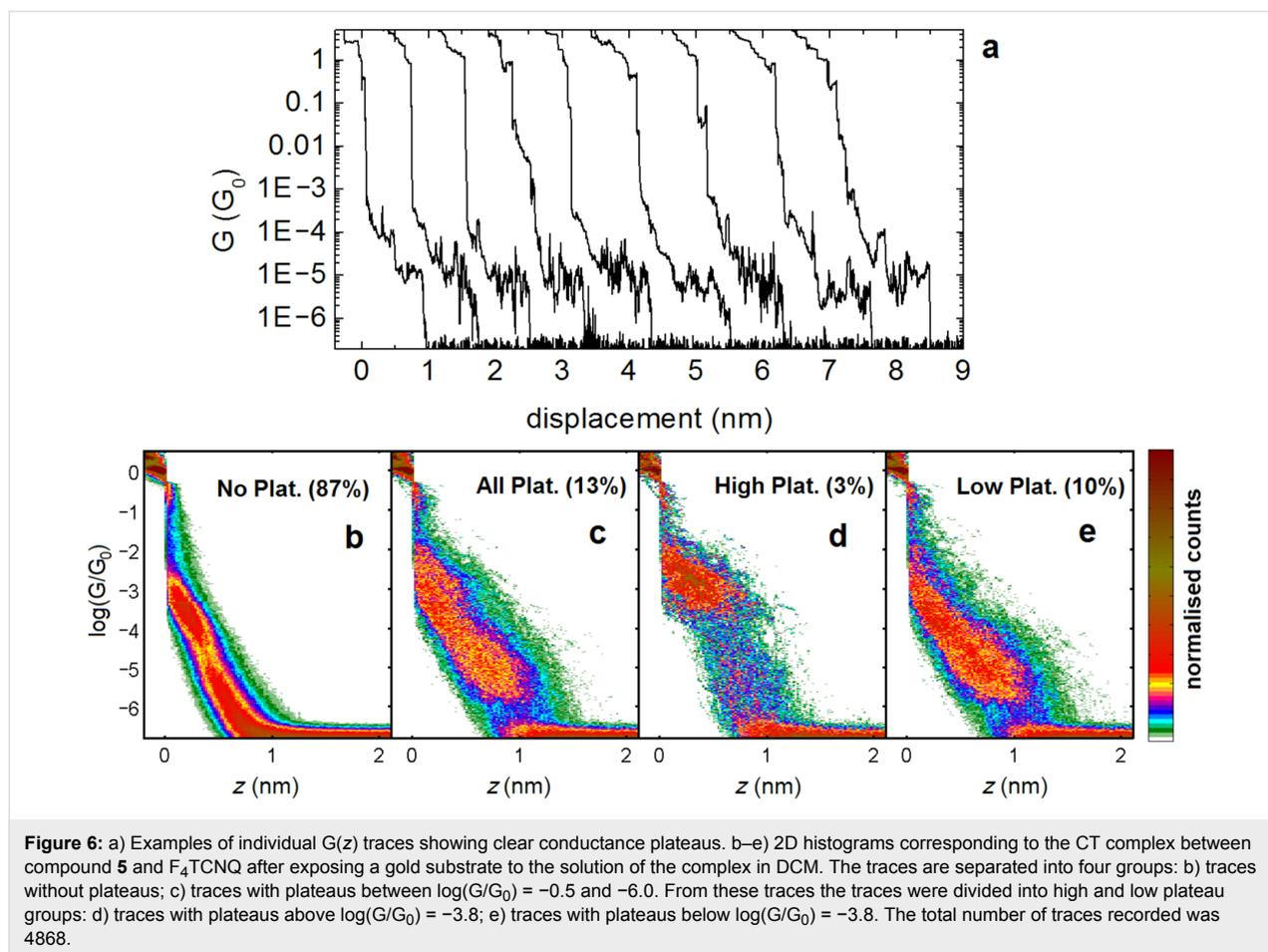

**Figure 6:** a) Examples of individual G(z) traces showing clear conductance plateaus. b–e) 2D histograms corresponding to the CT complex between compound **5** and $F_4TCNQ$ after exposing a gold substrate to the solution of the complex in DCM. The traces are separated into four groups: b) traces without plateaus; c) traces with plateaus between $\log(G/G_0) = -0.5$ and $-6.0$. From these traces the traces were divided into high and low plateau groups: d) traces with plateaus above $\log(G/G_0) = -3.8$; e) traces with plateaus below $\log(G/G_0) = -3.8$. The total number of traces recorded was 4868.

As can be seen from the separation of plateaus into high and low groups, the two types generally occur independently. The origin of the high conductance state is difficult to be totally sure about. The two groups may arise from independent chemical species, in which case it would be natural to label the high group as junctions for which transport takes place through a molecule of $F_4TCNQ$. The length and conductance are similar to the control test we carried out on only this molecule (see Supporting Information File 1, Figure S7). It may also be possible that the high conductance state arises through contact to the central part of the CT complex and one of the terminal thiol groups. We observed a similar feature for the neutral molecule (see Figure 4b and Figure 4c) although for the neutral molecule, this signal was never very reproducible. The low group, on the other hand, fits well with conductance taking place across the whole molecule (i.e., thiol to thiol). We calculate an S–S distance of 2.4 nm for the CT complex (substituting the dihydroanthracene core of compound **5** for anthracene). We estimate the amount of gold retraction to be 0.5 nm, which then gives a real Au–Au separation of $1.64 \pm 0.30$ nm for the mean breaking distance of the low group, corresponding well with the length of the molecule. The fact that the molecule does not seem to fully stretch inside the junctions (unlike the OPE3-dithiol) may be due to the bulky nature of the complex, which contains many groups with the potential to interact with gold.

We cannot be totally sure of the species which gives rise to the low conductance state in so much as we do not know its exact charge and ratio of donor to acceptor molecules. It is conceivable that some charge is transferred to the electrodes upon molecular junction formation, as is common for molecular junctions [33]. However, it is well known that the cationic (1+) state of exTTF is less thermodynamically stable than the dication (2+) [34], resulting in inverted oxidation potentials of the cation and dication. Additionally, it is known that the resonance between the neutral and dication is impossible due to the strong difference in geometries. Thus, it is more than plausible that the molecule retains its cationic state in the junction. We further point out that in several recent studies, a very similar conductance was found for a molecule that is structurally similar and contains only the anthracene central group. In a study by Hong et al., a conductance peak was found at $\log(G/G_0) = -4.5$, which is close to the low conductance group we observe [10]. As it is known that the oxidation of exTTF compounds results in the planarization and aromatization of the central three rings to the anthracene unit, the similarity in conductance is perhaps indicative of these changes. We have also recently shown that adding substituents to the central phenyl rings of OPE wires has no noticeable influence on the molecular conductance (groups were tested ranging from electron-withdrawing fluorine to electron-donating methoxy groups) [35]. It is also rational to suppose that the, albeit charged, dithiole side groups of our CT complex would not strongly modify the conductance of the otherwise neutral anthracene-containing wire. Finally, we point out that recent conductive AFM measurements carried out on self-assembled monolayers of anthracene and anthraquinone OPE derivatives showed a difference of approximately two orders of magnitude between their respective conductance values [36]. This would fit with our hypothesis that the conductance of the neutral exTTF molecule, which has a similar bonding pattern to anthraquinone, is below our experimental resolution, based on the value we measure for the CT complex.

## Ab initio calculations

To gain insight into the conduction mechanism of compound **5**, we performed theoretical calculations based on a combination of density functional theory (DFT) and Green's functions techniques within the framework of the Landauer theory for coherent transport. The complete technical details are reported in Supporting Information File 1.

We first optimized the geometry of the molecule in the gas phase. The HOMO was found to be localized on the exTTF unit, which presented the expected butterfly shape, while the LUMO appeared delocalized over the whole molecule (Figure 7).

Subsequently, we constructed two kinds of metal–molecule–metal junctions, where the molecule is bound to a gold cluster in a top and hollow position, respectively (Figure 8). We then computed the zero-bias electron transmission following the procedure explained in Supporting Information File 1. The corresponding transmission curves are shown in Figure 9. Notice that the molecular HOMO–LUMO gap was corrected, following the procedure previously reported [37]. The HOMO and all other occupied orbital energies were shifted by $\Sigma_{occ} = -IP - \varepsilon_H + \Delta_{occ}$, while the LUMO and all other unoccupied orbital levels were shifted by $\Sigma_{virt} = -EA - \varepsilon_L + \Delta_{virt}$. Here, $\Delta_{occ}$ ($\Delta_{virt}$) is the image charge correction for the occupied (unoccupied) states, $\varepsilon_H$ ($\varepsilon_L$) is the Kohn–Sham energy of the gas phase HOMO (LUMO), and IP(EA) is the gas phase ionization potential (electron affinity). All quantities are reported in Table 1 for both binding geometries.

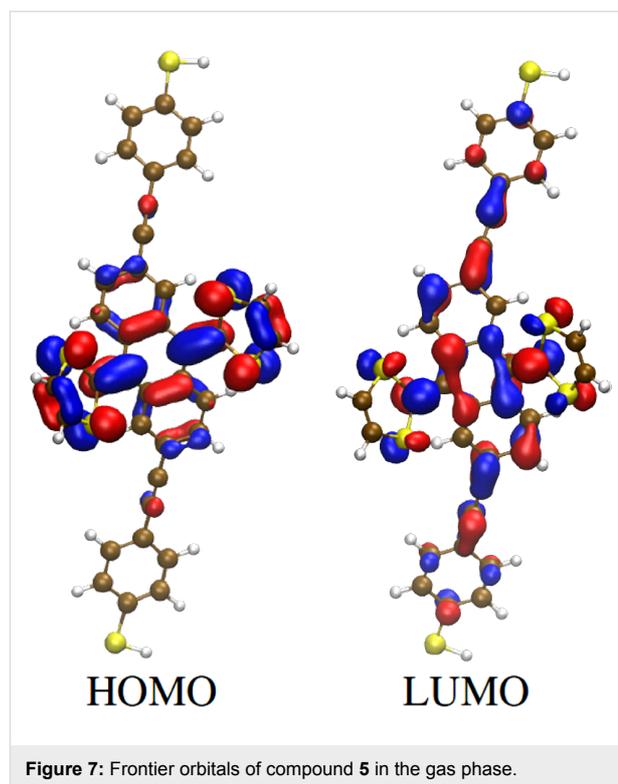

**Figure 7:** Frontier orbitals of compound **5** in the gas phase.

The alignment of the Breit–Wigner resonances related to both the HOMO and LUMO (at approximately −1 and 2.7 eV from the Fermi level, respectively) do not show a strong dependence on the binding geometry (Figure 9). The electron transport is dominated by the HOMO, although interference features (resonance–antiresonance pairs) appear in the energy region close to the Fermi level. This is not surprising, given the spatial extension of the frontier orbitals, in particular of the HOMO, which is localized on the ex-TTF unit only. In fact, Fano resonances are known to arise when a "pendant" orbital, which is weakly coupled to the electrodes, is coupled to an orbital delocalized over the main axis of the molecular wire [38,39].

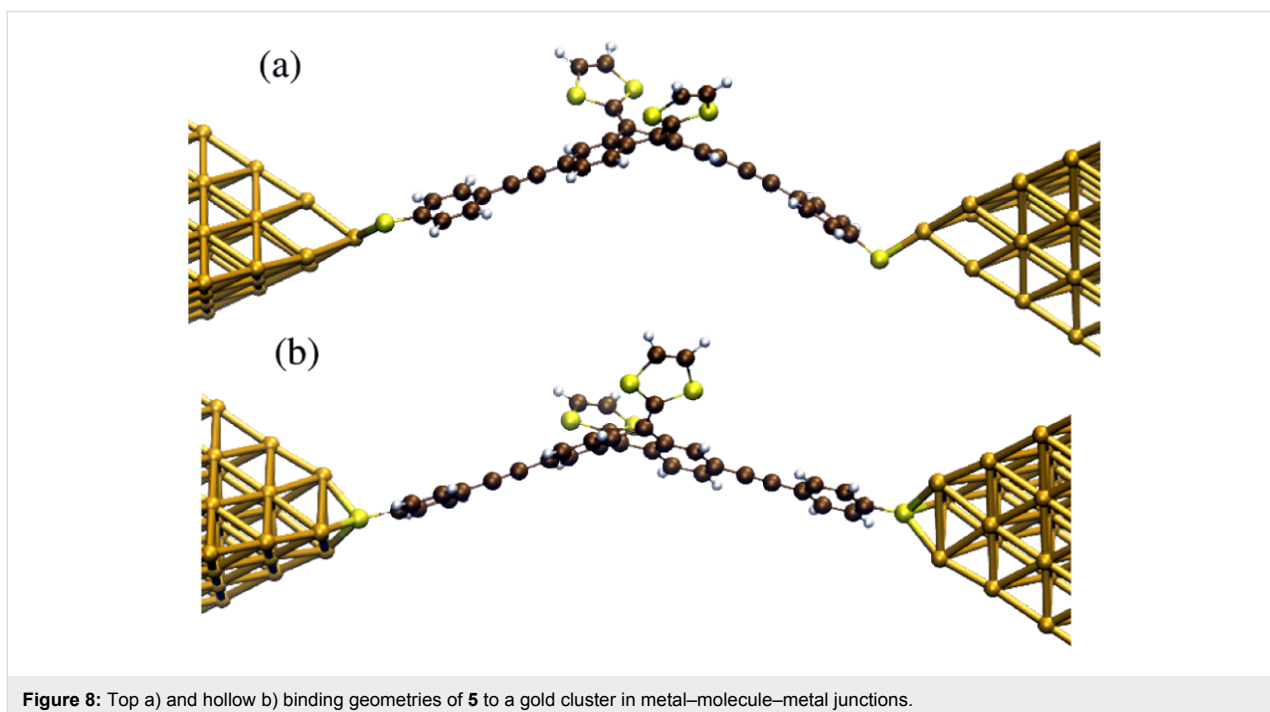

**Figure 8:** Top a) and hollow b) binding geometries of **5** to a gold cluster in metal–molecule–metal junctions.

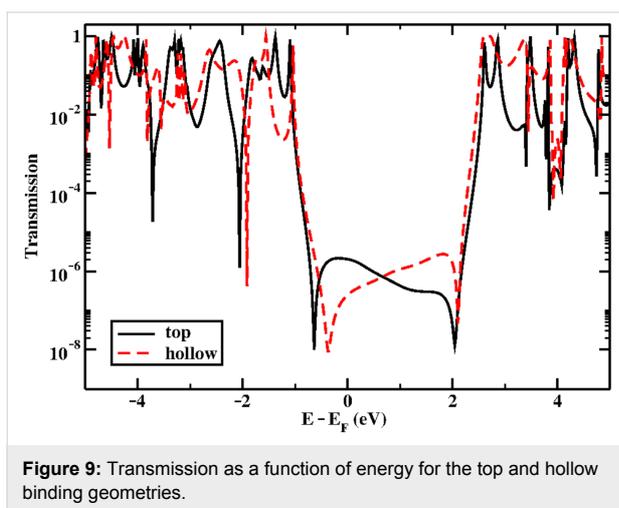

**Figure 9:** Transmission as a function of energy for the top and hollow binding geometries.

**Table 1:** Kohn–Sham HOMO and LUMO, ionization potential (IP), electron affinity (EA) and image charge correction for occupied $\Delta_{occ}$ and $\Delta_{virt}$ unoccupied orbitals. All quantities are in eV.

|        | HOMO  | IP   | $\Delta_{occ}$ | LUMO  | EA    | $\Delta_{virt}$ |
|--------|-------|------|----------------|-------|-------|-----------------|
| top    | −4.27 | 5.68 | −0.35          | −2.54 | −1.16 | −0.38           |
| hollow | −4.27 | 5.68 | −0.32          | −2.54 | −1.16 | −0.33           |

The computed conductance values, evaluated as the transmission at the Fermi level, are in the range $10^{-6}$ to $10^{-7}$ $G_0$. These low conductance values are at the limit of what we can observe experimentally. This, therefore, would be consistent with the idea that the real conductance of the molecule is too low to be recorded in the experiments. Nevertheless, we cannot absolutely rule out the possibility that the molecule does not form molecular junctions in the experiments.

## Conclusion

We have synthesized a molecular wire containing a π-extended tetrathiafulvalene (exTTF) group and studied its single-molecule electrical transport properties along with those of its charge-transfer complex with $F_4$TCNQ. Within the accessible conductance range (10 to $10^{-7}$ $G_0$) we did not observe a clear conductance signature of the neutral parent molecule. This alone could suggest either that its conductance is too low or that it does not form stable junctions. We did, however, find a clear conductance signature in the experiments carried out on the charge-transfer complex. As complexation with the acceptor oxidizes the molecule by removing two electrons from the exTTF group, thus converting it from a buckled and cross-conjugated group into a planar aromatic group, we predict the CT species to have a higher conductance than the neutral molecule. This, we believe, supports the idea that the conductance of the neutral molecule is very low, below our measurement sensitivity. This would make the conductance difference between the neutral and CT species at least two orders of magnitude. This can be considered as favorable for the use of single molecules as chemical sensors, in which analyte molecules may bind to a backbone to alter its conductance. Further combinations of donors and acceptors should be explored in order to evaluate this potential.

## Supporting Information

### Supporting Information File 1

Detailed experimental procedures for the synthesis and characterization of **5**, break junction experiments and theoretical methods.
[http://www.beilstein-journals.org/bjoc/content/supplementary/1860-5397-11-120-S1.pdf]

## Acknowledgements

We thank Dr. Jose Manuel Santos Barahona for help with preparing the CT complex with TCNQ, and recording and interpreting the UV–vis spectra. Financial support by the European Commission (EC) FP7 ITN "MOLESCO" Project No. 606728, the European Research Council (ERC-436 2012 ADG_20120216-Chirallcarbon), the CAM (PHOTOCARBON project S2013/MIT-2841, NANOFRONTMAG-CM project S2013/MIT-2850 and MAD2D project S2013/MIT-3007), FP7-ENERGY-2012-1-2STAGE-number 309223 (PHOCS) and the Spanish MICINN/MINECO through the programs MAT2011-25046, MAT2014-57915-R and PRI-PIBUS-2011-1067 is acknowledged. MB was partly supported by a FY2012 (P12501) Postdoctoral Fellowship for Foreign Researchers from the Japan Society for Promotion of Science (JSPS) and by a JSPS KAKENHI, "Grant-in-Aid for JSPS Fellows", grant no. 24·02501. YA is also thankful to another KAKENHI, "Grant-in-Aid for Scientific Research on Innovation Areas, Molecular Architectonics: Orchestration of Single Molecules for Novel Functions" (#25110009). LAZ was supported by the Spanish MICINN under Grant MAT2011-23627. FP acknowledges support by the Carl-Zeiss foundation and the Collaborative Research Center 767 "Controlled Nanosystems: Interaction and Interfacing to the Macroscale".

## References

1. Leary, E.; La Rosa, A.; González, M. T.; Rubio-Bollinger, G.; Agraït, N.; Martín, N. *Chem. Soc. Rev.* **2015,** *44,* 920–942. doi:10.1039/C4CS00264D
2. Guldi, D. M.; Nishihara, H.; Venkataraman, L., Eds. Molecular wires. *Chem. Soc. Rev.* **2015,** *4,* 835–1030. doi:10.1039/C5CS90010G
3. Forrest, S. R.; Thompson, M. E., Eds. Organic Electronics and Optoelectronics. *Chem. Rev.* **2007,** *107,* 923–1386. doi:10.1021/cr0501590
4. Guldi, D. M.; Illescas, B. M.; Atienza, C. M.; Wielopolski, M.; Martín, N. *Chem. Soc. Rev.* **2009,** *38,* 1587–1597. doi:10.1039/b900402p
5. Haiss, W.; van Zalinge, H.; Higgins, S. J.; Bethell, D.; Höbenreich, D.; Schiffrin, D. J.; Nichols, R. J. *J. Am. Chem. Soc.* **2003,** *125,* 15294–15295. doi:10.1021/ja038214e
6. Gittins, D. I.; Bethell, D.; Schiffrin, D. J.; Nichols, R. J. *Nature* **2000,** *408,* 67–69. doi:10.1038/35040518
7. Chen, F.; He, J.; Nuckolls, C.; Roberts, T.; Klare, J. E.; Lindsay, S. *Nano Lett.* **2005,** *5,* 503–506. doi:10.1021/nl0478474
8. Chen, F.; Nuckolls, C.; Lindsay, S. *Chem. Phys.* **2006,** *324,* 236–243. doi:10.1016/j.chemphys.2005.08.052
9. Xu, B. Q.; Li, X. L.; Xiao, X. Y.; Sakaguchi, H.; Tao, N. J. *Nano Lett.* **2005,** *5,* 1491–1495. doi:10.1021/nl050860j
10. Hong, W.; Valkenier, H.; Mészáros, G.; Manrique, D. Z.; Mishchenko, A.; Putz, A.; Moreno García, P.; Lambert, C.; Hummelen, J. C.; Wandlowski, T. *Beilstein J. Nanotechnol.* **2011,** *2,* 699–713. doi:10.3762/bjnano.2.76
11. Xiao, X.; Brune, D.; He, J.; Lindsay, S.; Gorman, C. B.; Tao, N. *Chem. Phys.* **2006,** *326,* 138–143. doi:10.1016/j.chemphys.2006.02.022
12. Taniguchi, M.; Tsutsui, M.; Shoji, K.; Fijiwara, H.; Kawai, T. *J. Am. Chem. Soc.* **2009,** *131,* 14146–14147. doi:10.1021/ja905248e
13. Giacalone, F.; Ángeles Herranz, M.; Grüter, L.; González, M. T.; Calame, M.; Schönenberger, C.; Arroyo, C.; Rubio-Bollinger, G.; Vélez, M.; Agraït, N.; Martín, N. *Chem. Commun.* **2007,** 4854–4856. doi:10.1039/b710739k
14. Li, M.-J.; Long, M.-Q.; Chen, K.-Q.; Xu, H. *Solid State Commun.* **2013,** *157,* 62–67. doi:10.1016/j.ssc.2012.12.001
15. Leary, E.; Higgins, S. J.; van Zalinge, H.; Haiss, W.; Nichols, R. J.; Nygaard, S.; Jeppesen, J. O.; Ulstrup, J. *J. Am. Chem. Soc.* **2008,** *130,* 12204–12205. doi:10.1021/ja8014605
16. Liao, J.; Agustsson, J. S.; Wu, S.; Schönenberger, C.; Calame, M.; Leroux, Y.; Mayor, M.; Jeannin, O.; Ran, Y.-F.; Liu, S.-X.; Decurtins, S. *Nano Lett.* **2010,** *10,* 759–764. doi:10.1021/nl902000e
17. Parker, C. R.; Leary, E.; Frisenda, R.; Wei, Z.; Jennum, K. S.; Glibstrup, E.; Abrahamsen, P. B.; Santella, M.; Christensen, M. A.; Della Pia, E. A.; Li, T.; González, M. T.; Jiang, X.; Morsing, T. J.; Rubio-Bollinger, G.; Laursen, B. W.; Nørgaard, K.; van der Zant, H.; Agraït, N.; Nielsen, M. B. *J. Am. Chem. Soc.* **2014,** *136,* 16497–16507. doi:10.1021/ja509937k
18. Martín, N.; Sánchez, L.; Herranz, M. A.; Illescas, B.; Guldi, D. M. *Acc. Chem. Res.* **2007,** *40,* 1015–1024. doi:10.1021/ar700026t
19. Brunetti, F. G.; López, J. L.; Atienza, C.; Martín, N. *J. Mater. Chem.* **2012,** *22,* 4188–4205. doi:10.1039/c2jm15710a
20. Giacalone, F.; Segura, J. L.; Martín, N.; Guldi, D. M. *J. Am. Chem. Soc.* **2004,** *126,* 5340–5341. doi:10.1021/ja0318333
21. Molina-Ontoria, A.; Wielopolski, M.; Gebhardt, J.; Gouloumis, A.; Clark, T.; Guldi, D. M.; Martín, N. *J. Am. Chem. Soc.* **2011,** *133,* 2370–2373. doi:10.1021/ja109745a
22. Wielopolski, M.; Molina-Ontoria, A.; Schubert, C.; Margraf, J. T.; Krokos, E.; Kirschner, J.; Gouloumis, A.; Clark, T.; Guldi, D. M.; Martín, N. *J. Am. Chem. Soc.* **2013,** *135,* 10372–10381. doi:10.1021/ja401239r
23. Illescas, B. M.; Santos, J.; Martín, N.; Atienza, C. M.; Guldi, D. M. *Eur. J. Org. Chem.* **2007,** 5027–5037. doi:10.1002/ejoc.200700226
24. Pearson, D. L.; Tour, J. M. *J. Org. Chem.* **1997,** *62,* 1376–1387. doi:10.1021/jo962335y
25. Liu, S.-G.; Pérez, I.; Martín, N.; Echegoyen, L. *J. Org. Chem.* **2000,** *65,* 9092–9102. doi:10.1021/jo001149w
26. Herranz, M. A.; Yu, L.; Martín, N.; Echegoyen, L. *J. Org. Chem.* **2003,** *68,* 8379–8385. doi:10.1021/jo034894s
27. González, M. T.; Leary, E.; García, R.; Verma, P.; Herranz, M. A.; Rubio-Bollinger, G.; Martín, N.; Agraït, N. *J. Phys. Chem. C* **2011,** *115,* 17973–17978. doi:10.1021/jp204005v
28. Chen, F.; Tao, N. J. *Acc. Chem. Res.* **2009,** *42,* 429–438. doi:10.1021/ar800199a

29. Urban, C.; Écija, D.; Wang, Y.; Trelka, M.; Preda, I.; Vollmer, A.; Lorente, N.; Arnau, A.; Alcamí, M.; Soriano, L.; Martín, N.; Martín, F.; Otero, R.; Gallego, J. M.; Miranda, R. *J. Phys. Chem. C* **2010,** *114,* 6503–6510. doi:10.1021/jp911839b
30. Jain, A.; Rao, K. V.; Mogera, U.; Sagade, A. A.; George, S. J. *Chem. – Eur. J.* **2011,** *17,* 12355–12361. doi:10.1002/chem.201101813
31. Bryce, M. R.; Moore, A. J.; Hasan, M.; Ashwell, G. J.; Fraser, A. T.; Clegg, W.; Hursthouse, M. B.; Karaulov, A. I. *Angew. Chem., Int. Ed. Engl.* **1990,** *29,* 1450–1452. doi:10.1002/anie.199014501
32. Bryce, M. R.; Finn, T.; Batsanov, A. S.; Kataky, R.; Howard, J. A. K.; Lyubchik, S. B. *Eur. J. Org. Chem.* **2000,** 1199–1205. doi:10.1002/1099-0690(200004)2000:7<1199::AID-EJOC1199>3.0.CO;2-F
33. Peng, G.; Strange, M.; Thygesen, K. S.; Mavrikakis, M. *J. Phys. Chem. C* **2009,** *113,* 20967–20973. doi:10.1021/jp9084603
34. Bendikov, M.; Wudl, F.; Perepichka, D. F. *Chem. Rev.* **2004,** *104,* 4891–4946. doi:10.1021/cr030666m
35. González, M. T.; Zhao, X.; Manrique, D. Z.; Miguel, D.; Leary, E.; Gulcur, M.; Batsanov, A. S.; Rubio-Bollinger, G.; Lambert, C. J.; Bryce, M. R.; Agraït, N. *J. Phys. Chem.* **2014,** *118,* 21655–21662. doi:10.1021/jp506078a
36. Guédon, C. M.; Valkenier, H.; Markussen, T.; Thygesen, K. S.; Hummelen, J. C.; van der Molen, S. J. *Nat. Nanotechnol.* **2012,** *7,* 305–309. doi:10.1038/nnano.2012.37
37. Zotti, L. A.; Bürkle, M.; Pauly, F.; Lee, W.; Kim, K.; Jeong, W.; Asai, Y.; Reddy, P.; Cuevas, J. C. *New J. Phys.* **2014,** *16,* 015004. doi:10.1088/1367-2630/16/1/015004
38. Cuevas, J. C.; Scheer, E. *Molecular electronics: An Introduction to Theory and Experiment;* World Scientific Publishing Co Pte Ltd: Singapore, 2010; Vol. 1. doi:10.1142/7434
39. Lambert, C. J. *Chem. Soc. Rev.* **2015,** *44,* 875–888. doi:10.1039/C4CS00203B